\documentstyle[aps,manuscript]{revtex}

\newcommand{\AmS}{{\protect\the\textfont2
  A\kern-.1667em\lower.5ex\hbox{M}\kern-.125emS}}

\title{An investigation of interplay between dissipation mechanisms in 
heated Fermi systems by means of radiative strength functions}

\author{V.A. Plujko (Plyuyko)\address{
Institute for Nuclear Research, Pr. Nauki 47, 252028, Kiev-28, Ukraine}}  

\begin{document}

\maketitle

\begin{abstract}

A simple analytical expression for the $\gamma$-decay strength function is 
derived with microcanonical ensemble for initial excited states.
The approach leads to both a non-zero limit of the strength function for
vanishing gamma-ray energy and a partial breakdown of  Brink hypothesis.
It is shown that the low energy behaviour of the  $\gamma$~- decay strength 
functions is governed by the energy behavior of the 
damping width. It may provides a new tool for study of the interplay between 
different relaxation mechanisms of the collective excitations.

\end{abstract}

\section{INTRODUCTION}

 The emission and absorption of the gamma-rays as well as  electron~- 
positron decay are described in the many-body systems  
by the radiative  strength functions. 
These functions are very useful for study  nuclear models, mechanisms of 
the $\gamma$- processes, widths of the collective excitations and nuclear 
deformations \cite{SK1986,AY1988,g92,Bracco95,brt,mbc}. Besides fundamental 
importance from a theoretical point of view, the strength functions 
are needed to generate data for the energy and non- energy applications 
\cite{RIPL}. It is critycally important to have the simple 
closed-form expression for the $\gamma$~- ray strength function because  
in the most cases this function is auxiliary quantity used in calculations 
of different nuclear characteristics and processes. 

\section{ GAMMA-DECAY STRENGTH FUNCTION IN HEATED NUCLEI}

The gamma- decay strength function 
$\overleftarrow{f}_{E1}(\epsilon_{\gamma})$ determines the $\gamma$-emission 
from heated nuclei~\cite{Bart73}. It is connected with the average 
$\gamma$-width $\bar{\Gamma}(\epsilon_{\gamma })$  for radiation of the energy 
$\epsilon_{\gamma}$ as 
$
\bar{\Gamma}(\epsilon_{\gamma })  =   
 3 \epsilon_{\gamma}^{3} \overleftarrow{f}_{E1}(\epsilon_{\gamma})
\Omega_{f} (E_{f}=E-\epsilon_{\gamma})/ 
\Omega_{i} (E) ,
$
where  $\Omega_{i}$ and $\Omega_{f}$ are the total density of the initial
and final states, respectively. The dipole transitions is considered for 
reasons of  importance in applications. We  find the expression for the 
function $\overleftarrow{f}_{E1}$ by use of the relation for average 
radiative width from~\cite{plu90}. The latter expression was obtained
with the microcanonical distribution for excited states. We have
in the case of the spherical nuclei
\begin{equation}
\overleftarrow{f}_{E1}(\epsilon_{\gamma})   
= - 6 e^{2} 
{ N Z  \over A }  \left({2\over 3\hbar c }\right)^3 
{Im\chi^{(-)}(\omega,T_{f}) \over 
1-\exp (-\epsilon_{\gamma } / T_{f})} ,
\label{m1}
\end{equation}
\noindent where 
$\chi^{(-)}(\omega, T_{f}) =
Sp(r Y_{1 0 }(\hat{r}) \delta n^{(-)})/ q_{\omega}(t)$ 
is the linear response function of the heated nucleus to the external 
dipole field $q_{\omega}(t)~r~Y_{1 0 }(\hat{r})$, 
$q_{\omega}(t)=~q_{0}\exp[-i(\omega+i\eta)t]$, $\eta \to +0$, $q_{0} \ll 1$,
and $\delta n^{(-)}(t) \equiv \delta n_{p}(t) -\delta n_{n}(t)$ is the 
perturbation of the isovector single-particle density induced by this field.
The function $\chi^{(-)}$ is proportional to the polarizability
of the nucleus in the electric dipole field. 

Note that the $\gamma$- decay strength function  depends on 
temperature $T_{f}$ of the final states. This temperature is a function 
of the $\gamma$- ray energy in contrast to the initial states temperature $T$. 

We use the hydrodynamic model with the friction~\cite{eisgre} to 
provide a simple closed-form expression for the response function. 
This approach is the extention of the 
Steinwedel- Jensen (SJ) hydrodynamic model and gives a simple description 
of the giant dipole resonance (GDR) excitation as well as its 
damping. The standart hydrodynamics corresponds to the Vlasov-Landau kinetic 
approach when only the monopole and dipole distortions of the Fermi sphere 
are taken into account~\cite{yh81}. Because of this, we apply the 
Vlasov-Landau kinetic equation completed by a source term for relaxation 
processes in order to obtain the friction coefficient of the isovector 
velocity. The SJ-mode plays the most important role in heavy nuclei 
\cite{msk77}. It corresponds to a volume density oscillation which 
is almost unaffected by the dynamical distortion of the Fermi surface 
with  multipolarities more than quadrupole \cite{na84}. 
Next we follow the approach from \cite{plu98} and do not use a 
normalization of the damping width to that magnitude which corresponds to 
the infinite matter value \cite{KPS2,plu97}. 

 One finally gets for the damping width of the  isovector velocity
\begin{equation}
\Gamma_{\Gamma}(\epsilon_{\gamma}, T_{f}) = 
\Gamma_{c}(\epsilon_{\gamma},T_{f}) + \Gamma_{s}, \ \ \
\Gamma_{c}(\epsilon_{\gamma},T_{f}) = \hbar / \tau_{c}
(\epsilon_{\gamma},T_{f}), \ \ \Gamma_{s} = k_{s}\hbar  \bar{v} \ R_{0}.
\label {w8}
\end{equation} 
\noindent Here, $\Gamma_{c}$ and $\Gamma_{s}$ are the two-body and 
one-body (fragmentation) contributions to the total width respectively. The 
quantity $\tau_{c}(\epsilon_{\gamma},T_{f})$ is the collisional relaxation 
time for the isovector dipole distortion of the Fermi surface. It is 
associated with two-body collisions in the heated nucleus which is subjected 
to the electric  field oscillating with the frequency 
$\omega = \epsilon_{\gamma}/\hbar$. For the 
isotropic collision probabilities it is given by \cite{plu98,KPS2,ay}
\begin{equation}
\tau_{c}(\epsilon_{\gamma},T_{f}) \simeq 0.9 \hbar \alpha  T_{f}^{-2} 
/ [1+ ( \epsilon_{\gamma} /2\pi T_{f})^{2}].
\label {tauc}
\end{equation}
\noindent The dependence of the relaxation time $\tau_{c}$ on the energy 
$\epsilon_{\gamma}$ results from memory effects in the collision integral 
and follows Landau's prescription. The temperature dependence arises from 
the smeared out behavior of the equilibrium distribution function near the 
Fermi momentum in the heated nuclei. The one-body relaxation width 
$\Gamma_{c}$  is taken similar to the wall formula \cite{yan2,msk77} but 
with scaled coefficient $k_{s}$ \cite{KPS2}. The quantities $R_{0}$ 
and $\bar{v} \approx (3v_F/4)$   are  the nuclear radius and the average 
velocity of the nucleon, respectively. 

Using the expression for the polarizability of the nucleus
in the dipole mode approximation from \cite{eisgre} and Eq.(\ref{m1}) 
we get for the dipole strength function
\begin{equation}
\overleftarrow{f}_{E1}(\epsilon_{\gamma}) =  
8.674 \cdot 10^{-8} 
{ \sigma_{0} \Gamma_{G} \over 1-\exp (-\epsilon_{\gamma } / T_{f})} 
{\epsilon_{\gamma} \Gamma_{\gamma}(\epsilon_{\gamma},T_{f}) \over
 ( \epsilon_{\gamma}^2 -  E_{G}^2)^2 +
(\Gamma_{\gamma}(\epsilon_{\gamma},T_{f}) \epsilon_{\gamma})^2} ,
\ \ \  (~MeV^{-3}~),
\label{fe6} 
\end{equation} 
\noindent where $E_{G}$ and $\Gamma_{G}$ are  the GDR energy and  
width, respectively, in $MeV$; the quantity $\sigma_{0}$ is the peak of the 
photoabsorption cross-section in $mb$.
 
This approach takes into account various damping mechanisms as well as 
thermal energy of the electromagnetic field in heated nuclei. The imaginary 
part of the dipole response function associated with  Eq.(\ref{fe6}) has 
a Lorentzian shape with frequency-dependent width. In the cold nuclei this 
form of the $Im\chi^{(-)}$ was obtained within the random-phase 
approximation~\cite{DLH1972}. This term is also in close agreement with 
the imaginary part of the response function of the heated 
Fermi~- liquid drop  to an external pressure, when approximation of
the dissipative nuclear fluid-dynamics is used for description of the 
system~\cite{plu97}.

\section{NUMERICAL RESULTS AND DISCUSSIONS}

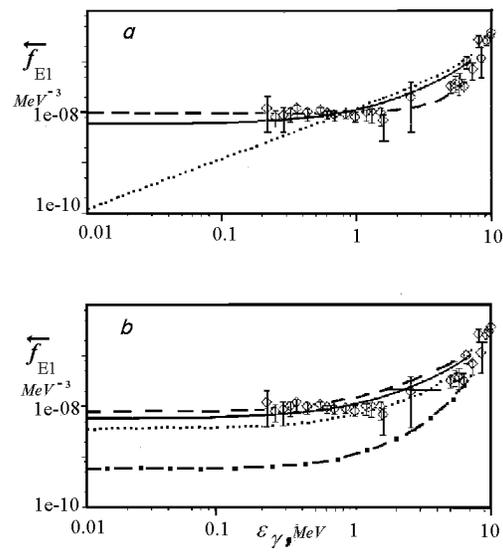
In Fig.1 results of the calculations of the strength functions 
$\overleftarrow{f}_{E1}$ in ${}^{144}Nd$ with the  initial states energy 
$E$ equal to the neutron binding energy $B_{n}=7.82~MeV$ are shown. 
The GDR parameters were taken from photonuclear data \cite{db}. We used 
the Fermi gas model to get the temperature $T_{f}$ of the final states, 
$
T_{f} = \sqrt{T^{2} - \epsilon_{\gamma} / a} , E = a T^{2},
$
where $a$ is the level density parameter, $a=A/8$. 
The value of the scaled coefficient $k_{s}$ was found to fit the quantity 
$\Gamma_{\gamma}(\epsilon_{\gamma}=E_{G}, T=0)$ to the GDR width 
$\Gamma_{G}$. The experimental data are taken from \cite{expNd}.
All solid curves were obtained by  use of the  Eq.(\ref{fe6}) with 
$\alpha = 9.2 MeV$. This value $\alpha$ corresponds to the magnitude 
of the in-medium nucleon-nucleon cross section which is smaller than the 
cross section in free space by a factor of 2 (see~\cite{KPS2} 
for comments). The dashed line in the Fig.1a was calculated 
within the framework of the EGLO model as presented in~\cite{RIPL,kuc93}:
\begin{equation}
\overleftarrow{f}_{E1}=
 8.674 \cdot 10^{-8}  \sigma_{0} \Gamma_{G} \left[ 
{\epsilon_{\gamma} \Gamma_{EL}(\epsilon_{\gamma},T) \over
 ( \epsilon_{\gamma}^2 -  E_{G}^2)^2 +
(\Gamma_{EL}(\epsilon_{\gamma},T) \epsilon_{\gamma})^2} +
0.7 {\Gamma_{EL}(\epsilon_{\gamma} = 0, T) \over E_{G}^{3}}
\right] ,
\label{fe10}
\end{equation}
\noindent where $\Gamma_{EL}(\epsilon_{\gamma},T) = [ k_{0} + (1 - k_{0})
(\epsilon_{\gamma} - \epsilon_{0}^{\gamma}) / 
( E_{G} - \epsilon_{0}^{\gamma})$ $\Gamma_{c 0}(\epsilon_{\gamma},T)$
is the empirical width, and $\Gamma_{c 0}(\epsilon_{\gamma},T)
\equiv (\epsilon_{\gamma}^{2} + (2\pi T)^{2}) \Gamma_{G} / E_{G}^{2}$
is the collisional width reproducing the GDR width data at
$\epsilon_{\gamma} = E_{G}$ and $T = 0$. The parameters 
$k_{0}$ and $\epsilon_{0}^{\gamma} = 4.5 \,MeV$ are adjusted to 
reproduce the averaged resonance capture data; 
$k_{0} = 1$ for the ${}^{144}Nd$. 

The dotted line presents the approximation of the $\gamma$- decay strength 
function  by  the Lorentzian  with the energy independent width 
$\Gamma_{G}$ (SLO model). In the Fig.1b the dot-dashed and 
dotted lines  show the calculations when damping width 
$\Gamma_{\gamma}$ is determined only  by the two- body and 
one- body dissipation mechanisms, respectively. The curve 3 
gives the evaluations with the quantity $\alpha = 4.6 \,MeV$
which corresponds to the nucleon-nucleon cross section in free space.

The results obtained by our approach and EGLO model are in good agreement  
at low energies.  They describe experimental data in this range much better 
than the SLO model and give  a nonzero temperature-dependent limit of the 
strength function for vanishing gamma-ray energy. 

When compared to EGLO approach in the range of the GDR peak energy, 
the behaviour of the E1 strength functions calculated by the proposed method 
is almost in coincidence with SLO model. It is resulted mainly of taking 
into accountf the one-body component of width which is practically independent 
of the gamma-ray energy. The EGLO model also includes an additional 
term; it is second component on the right-hand side of Eq.(\ref{fe10}).
Recall that the SLO approach is probably the most appropriate simple method 
for the estimation of the $\gamma$-strength in the range of giant resonance 
peak energy.

As seen it is from this figure, the values $\overleftarrow{f}_{E1}$ are 
sensitive to the magnitude of $\alpha$ which defines the 
contributions one-body and two-body components to  the GDR damping width.
A rather good description of experimantal data is obtained at 
$\alpha \approx 9.2 \,MeV$. In this case the contribution of the  collisional 
damping to the GDR width is about $15\%$; the latter is in agreement with 
the results of the direct fitting of experimental data for the GDR 
widths \cite{plu98}.

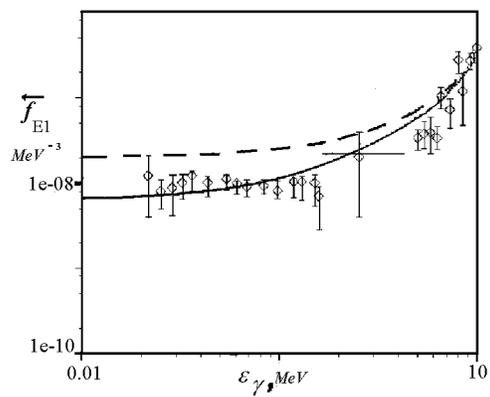
Figure 2 demonstrates the dependence of the dipole $\gamma$~-decay strength 
functions on the excitation energy $E$, i.e. a partial breakdown of Brink 
hypothesis. For the solid curve $E = B_{n}$ and $E = 50~MeV$ in the case 
of the dashed line. The violation 
of Brink hypotesis is growing with increasing  excitation energy. The 
difference of the  E1 strength function values calculated at different $E$ 
increases with decreasing $\gamma$~- energies and these deviations are more 
important for the $\gamma$~- transitions with energies under or of the order
of the nuclear temperature. 

\vspace{0.33cm}

I am very grateful to  Profs. A. Bracco and P.F. Bortignon 
for the kind hospitality at the GR98 Conference.
This work is supported in part by the International Atomic Energy
Agency (Vienna) under contract No.10308/RO.

\end{document}